\documentclass{PoS}
\usepackage{epsfig}

\newcommand{\lsim}{\raisebox{-0.7ex}{$\stackrel{\textstyle <}{\sim}$ }}

\title{Strong-Isospin Breaking in the 
Neutron-Proton Mass Difference}

\ShortTitle{Neutron-Proton Mass Difference}

\author{\speaker{Martin J.~Savage}\thanks{I would like to thank my
    collaborators on  this work, Silas Beane and Kostas Orginos. \qquad NT@UW-06-20.
}\\
        Department of Physics, University of Washington, 
Seattle, WA 98195-1560.\\
        E-mail: \email{savage@phys.washington.edu}}

\abstract{We determine the strong-isospin violating component of the
  neutron-proton mass difference from 
fully-dynamical lattice QCD and partially-quenched QCD calculations of the
nucleon mass, 
constrained by partially-quenched chiral perturbation theory at one-loop
level. 
The lattice calculations were performed with domain-wall valence quarks on MILC
lattices with rooted staggered sea-quarks at a lattice spacing of 
$b\sim0.125~{\rm fm}$, 
lattice spatial size of 
$L\sim2.5~{\rm fm}$ and pion masses ranging from 
$m_\pi \sim 290~{\rm MeV}$ to $\sim 350~{\rm MeV}$. 
At the physical value of the pion mass, we predict 
$M_n - M_p |^{(d-u)} = 2.26 \pm 0.57 \pm 0.42 \pm 0.10~{\rm MeV} $
where the first error is statistical, the second error is due to the
uncertainty in 
the ratio of light-quark masses, 
$\eta=m_u/m_d$, determined by MILC~\cite{Aubin:2004fs}, and the third
error 
is an estimate of the systematic due to chiral extrapolation. 
}

\FullConference{XXIV International Symposium on Lattice Field Theory\\
                 July 23-28 2006\\
                 Tucson Arizona, US}

\begin{document}

\section{Introduction}

It is a basic property of our universe that the neutron is
slightly more massive than the proton.  The electroweak interactions
are responsible for this mass difference, which receives contributions
from two sources.  
The strong-isospin breaking contribution (also
known as charge-symmetry breaking, for a review see
Ref.~\cite{Miller:2006tv}) is due to the difference in the masses of
the up and down quarks, ultimately determined by the values of the
Yukawa couplings in the Standard Model of electroweak interactions and
the vacuum expectation value of the Higgs field. 
With the down-quark more massive than the up-quark, the neutron would be more
massive than the proton in the absence of electromagnetism.
The experimental neutron-proton
mass difference of $M_n-M_p = 1.2933317\pm 0.0000005~{\rm
MeV}$~\cite{Eidelman:2004wy} receives an estimated electromagnetic
contribution of~\cite{Gasser:1982ap} 
$M_n-M_p\big|^{\rm em} = -0.76\pm 0.30~{\rm MeV}$ (estimated using the Cottingham sum-rule
saturated by the Born diagrams), 
and the remaining mass difference 
of $M_n-M_p\big|^{d-u}= 2.05\mp
0.30~{\rm MeV}$ is due to strong-isospin breaking.
The uncertainty is estimated from the Born
contribution of baryon-resonances to the Cottingham sum-rule.

Before delving into the details of the lattice calculation we have performed,
it is useful to remind ourselves why we should care about this mass-splitting in
the first place.
If we were in a situation where $M_n < M_p + M_e - M_\nu$, hydrogen would not be
stable.
It would decay weakly via $p+e\rightarrow n+\nu$ and chemistry of our universe
would be quite different from what we are familiar with.  
Therefore, in the limit of strong-isospin
symmetry, hydrogen would not be stable, and from our present calculation we find that
for hydrogen to be stable, $m_u/m_d \lsim 0.77$ (taken from the central values
of our determination and the electromagnetic contribution). 
Of course, one would still have some form of nuclear physics, but it might
look quite different from what we are familiar with.  In order to determine
just how similar, or different, nuclear physics would be, we need to perform
further calculations.

\section{The chiral expansion of the Nucleon Mass}

In order to extract  isospin breaking quantities from lattice calculations on
isospin symmetric lattices, one must perform partially-quenched calculations,
and then determine
coefficients in the chiral theory describing QCD quantities.
In QCD the proton mass has a chiral expansion of the form
\begin{eqnarray}
M_p & = & M_0 + \left(\overline{\alpha} + \overline{\beta} + 2 \overline{\sigma}\right) m_\pi^2 
\ -\ 
{1\over 3}\left(2 \overline{\alpha} - \overline{\beta}\right)\left({1 - \eta\over 1+\eta}\right)\ 
m_\pi^2\  
\nonumber\\
& - &
{1\over 8\pi f^2}\left[\ {3\over 2} g_A^2 m_\pi^3
\ +\ {4 g_{\Delta N}^2\over 3\pi} F_\pi\ \right]
\ \ \ ,
\label{eq:protmasspion}
\end{eqnarray}
where $\eta=m_u/m_d$ is the ratio of the mass of the up-quark and down-quark.
The function $F_\pi$ is given by 
$F_\pi = F(m_\pi,\Delta,\mu) $, with
\begin{eqnarray}
 F(m,\Delta,\mu) & = & 
\left(m^2-\Delta^2\right)\left(
\sqrt{\Delta^2-m^2} \log\left({\Delta -\sqrt{\Delta^2-m^2+i\epsilon}\over
\Delta +\sqrt{\Delta^2-m^2+i\epsilon}}\right)
-\Delta \log\left({m^2\over\mu^2}\right)\ \right)
\nonumber\\
& - & 
{1\over 2} m^2 \Delta  \log\left({m^2\over\mu^2}\right)
\ \ \ ,
\label{eq:massfun}
\end{eqnarray}
and arises from $\Delta$-resonance intermediate states in one-loop diagrams.
$\Delta$ is the mass difference between the $\Delta$-resonance and the nucleon.
The mass-splitting between the neutron and proton due to the quark mass
differences is therefore,
\begin{eqnarray}
M_n \ -\  M_p\big|^{d-u} & = & 
  {2\over 3}\left(2 \overline{\alpha} - \overline{\beta}\right)
\left({1 - \eta\over 1+\eta}\right)\ 
m_\pi^2\
\ \ \ .
\label{eq:mnminusmp}
\end{eqnarray}
The one-loop contributions at ${\cal O}(m_q^{3/2})$ cancel in the mass-difference,
as the pions are degenerate up to ${\cal O}(m_q^2)$.
Consequently, at leading order, it is the combination of parameters $2
\overline{\alpha} - \overline{\beta}$ that needs to be extracted from lattice
calculations.

The nucleon mass was computed in $SU(2)_L\otimes SU(2)_R$
partially-quenched chiral perturbation theory
(PQ$\chi$PT) in Ref.~\cite{Beane:2002vq}, and involves quite lengthy
expressions.
In the case of an isospin-symmetric set of lattice configurations, the proton
masses for different combinations of valence quarks can be found in
Ref.~\cite{Beane:2006fk}.
As an example, the mass of a proton with valence quantum numbers $V_1,V_1,V_2$
on a sea of $V_1$ quarks has the form
\begin{eqnarray}
& & M_p(V_1,V_1,V_2;V_1)\ =\  M_0 \ +\  2 \ \overline{\sigma}\
m^2_{V_1,V_1;V_1}    
\ +\
{1\over 6}\left(5\overline{\alpha} \ +\  2\overline{\beta}\right)\ m^2_{V_1,V_1;V_1}
\nonumber\\
&& 
\qquad
\ +\
{1\over 6}\left(\overline{\alpha} \ +\  4\overline{\beta}\right)\  m^2_{V_2,V_2;V_1}
\nonumber\\
& & 
\qquad
\ -\ 
{g_A^2\over 24\pi f^2} \left( {7\over 2} m^3_{V_1,V_1;V_1} \ +\
  m^3_{V_1,V_2;V_1} \right)
\nonumber\\
& & 
\qquad
\ -\ 
{g_A g_1\over 24\pi f^2} \left( {5\over 2} m^3_{V_1,V_1;V_1}  \ -\
  m^3_{V_1,V_2;V_1} 
\ -\ {3\over 2} m^3_{V_2,V_2;V_1}
\right)
\nonumber\\
& & 
\qquad
\ -\ 
{g_1^2\over 384\pi f^2} \left( 14 m^3_{V_1,V_1;V_1} 
\ +\  4 m^3_{V_1,V_2;V_1} 
\ -\ 27 m^3_{V_2,V_2;V_1}
\ +\ 9 m^2_{V_1,V_1;V_1} m_{V_2,V_2;V_1}
\right)
\\
& & 
\qquad
\ -\ 
{g_{\Delta N}^2\over 72\pi^2 f^2} 
\left(\ 2 F_{V_1,V_1;V_1}  + 9 F_{V_1,V_2;V_1} + F_{V_2,V_2;V_1} + 
m^2_{V_1,V_1;V_1} S_{V_2,V_2;V_1}
 - 
m^2_{V_2,V_2;V_1} S_{V_2,V_2;V_1}
\right);
\nonumber
\ \ ,
\label{eq:pqmassNPI}
\end{eqnarray}
where the function $S$ is given by 
$S_\pi = S(m_\pi,\Delta,\mu)$ with
\begin{eqnarray}
S(m,\Delta,\mu) & = & 
\sqrt{\Delta^2-m^2} 
\log\left({\Delta -\sqrt{\Delta^2-m^2+i\epsilon}\over
\Delta +\sqrt{\Delta^2-m^2+i\epsilon}}\right)
-\Delta \left( \log\left({m^2\over\mu^2}\right) + {1\over 3} \right)
\ \ \ .
\label{eq:massfunS}
\end{eqnarray}
$m_{V_a,V_b;V_c}$ is the mass of a pion composed of valence quarks $V_a, V_b$
on configurations with quarks $V_c$.
In partially-quenched proton masses, there are contributions that depend upon the
axial-isosinglet coupling, $g_1$.  This coupling is one of the fit
parameters in our extraction.

\section{The Lattice QCD Calculation}

\begin{table}[ht]
\vskip 0.1in
\begin{tabular}{|c|c|c|c|c|c|}
\hline
\label{table:configs}
 Ensemble        
& \ Theory \  
&\quad  $b m_l$ \quad 
&\quad $b m_{dwf}$\quad  
&\quad $10^3 \times b m_{res}$\quad 
& \# props.  \\
       \hline 
2064f21b679m007m050 & QCD&  0.007 ($V_1$) & 0.0081 ($V_1$)     & $1.604\pm 0.038$        & 468$\times$3 \\
2064f21b679m007m050 & PQ& 0.007 ($V_1$)& 0.0138 ($V_2$)    & $1.604\pm 0.038$        & 367$\times$3 \\
2064f21b679m007m050 & PQ& 0.007 ($V_1$)& 0.0100 ($V_3$)     & $1.604\pm 0.038$        & 367$\times$2 \\
2064f21b679m010m050 & QCD& 0.010 ($V_2$)& 0.0138 ($V_2$)    & $1.552\pm 0.027$       & 658$\times$3 \\
2064f21b679m010m050 & PQ& 0.010 ($V_2$)& 0.0081 ($V_1$)    & $1.552\pm 0.027$       & 658$\times$1 \\
  \hline
  \end{tabular}
  \caption{The parameters of the MILC gauge configurations and 
domain-wall propagators used in this work. 
For each propagator the extent of the fifth dimension is $L_5=16$.
The notation of quarks, $V_1, V_2, V_3$, is defined in the text.
The last column is the number of propagators generated, and corresponds to the
number of lattices times the number of different locations of sources on each
lattice.
We have worked at one lattice-spacing only, $b\sim 0.125~{\rm fm}$.
PQ denotes partially-quenched QCD.
For each set of configurations $b m_s = 0.050$.
}
\end{table}
\noindent 
Our computation~\cite{Beane:2006fk} uses the mixed-action lattice QCD scheme developed by
LHPC~\cite{Renner:2004ck,Edwards:2005kw} using domain-wall valence
quarks from a smeared-source on $N_f=2+1$
asqtad-improved~\cite{Orginos:1999cr,Orginos:1998ue} MILC
configurations generated with rooted~\footnote{For recent discussions
of the ``legality'' of the mixed-action and rooting procedures, see
Ref.~\cite{Durr:2004ta,Creutz:2006ys,Bernard:2006zw,Bernard:2006vv,Durr:2006ze,Hasenfratz:2006nw,Shamir:2006nj}.
While it still could be the case that staggered fermions do not reproduce QCD
in the continuum limit, recent work by Shamir~\cite{Shamir:2006nj}
on this
issue, relying heavily on the renormalization group evolution of the staggered theory, 
has concluded that under
certain plausible assumptions that can be tested, 
the continuum limit of the staggered theory is QCD.
}
staggered sea quarks~\cite{Bernard:2001av} that are hypercubic-smeared
(HYP-smeared)~\cite{Hasenfratz:2001hp,DeGrand:2002vu,DeGrand:2003in,Durr:2004as}. In
the generation of the MILC configurations, the strange-quark mass was
fixed near its physical value, $b m_s = 0.050$, (where $b\sim 0.125~{\rm
fm}$ is the lattice spacing) determined by the mass of hadrons
containing strange quarks.  
The two light quarks in the configurations
are degenerate (isospin-symmetric).  The domain-wall height is $m=1.7$
and the extent of the extra dimension is $L_5=16$.  The MILC lattices
were ``chopped'' using a Dirichlet boundary condition from 64 to 32
time-slices to save time in propagator generation.  In order to
extract the terms in the mass expansion, we computed a number of sets
of propagators corresponding to different valence quark masses, as
shown in Table~I.

On 468 $b m_l=0.007$ (denoted by $V_1$) lattices we have computed
three sets corresponding to the QCD point with a valence-quark mass of
$b m_{dwf}=0.0081$ ($V_1$), three sets on 367 $b m_l=0.007$ lattices
with a valence quark mass of $b m_{dwf}=0.0138$ (denoted by $V_2$),
and two sets with a valence quark mass $b m_{dwf}=0.0100$ (denoted by
$V_3$). On 658 of the $b m_l=0.010$ ($V_2$) lattices we have computed
three sets at the QCD point with a valence-quark mass of $b
m_{dwf}=0.0138$ ($V_2$) and one set with a valence quark mass of $b
m_{dwf}=0.0081$ ($V_1$). The parameters used to generate the QCD-point
light-quark propagators have been ``matched'' to those used to
generate the MILC configurations so that the mass of the pion computed
with the domain-wall propagators is equal (to few-percent precision)
to that of the lightest staggered pion computed with the same
parameters as the gauge configurations~\cite{Bernard:2001av}.  The
lattice calculations were performed with the {\it Chroma} software
suite~\cite{Edwards:2004sx,sse2} on the high-performance computing
systems at the Jefferson Laboratory (JLab).
\begin{figure}[!ht]
\vskip 0.35in
\centerline{{\epsfxsize=5.0in \epsfbox{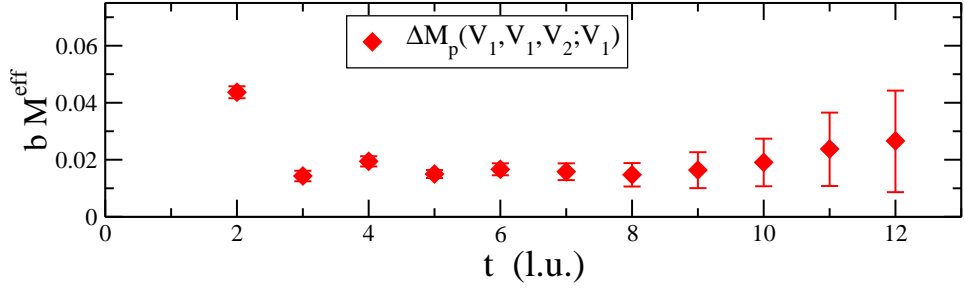}}} 
\noindent
\vskip 0.15in
\caption{\it 
The effective mass plot for the mass splitting $\Delta M_p(V_1,V_1,V_2;V_1)$.
}
\label{fig:DELTAMpeffmassA}
\vskip .2in
\end{figure}
To extract the isospin-breaking terms it is sufficient to look at
mass-differences between proton states.  The ratio of correlation
functions is formed so that a mass-difference can be extracted from its
large-time behavior. An example of the ratio of correlation functions found in
our work is shown in fig.~\ref{fig:DELTAMpeffmassA}.
We define the mass difference to be 
\begin{eqnarray}
\Delta M_p(V_a,V_b,V_c;V_d)& =  & M_p(V_a,V_b,V_c;V_d)\ -\ 
M_p(V_d,V_d,V_d;V_d)
\ \ \ .
\end{eqnarray}

\begin{figure}[!ht]
\vskip 0.35in
\centerline{{\epsfxsize=4.5in \epsfbox{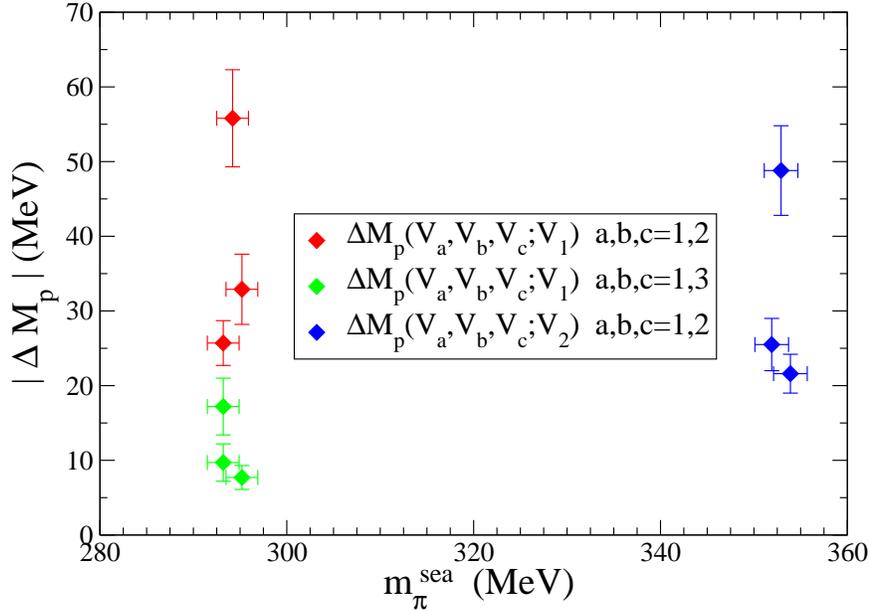}}} 
\noindent
\vskip 0.15in
\caption{\it 
The partially-quenched proton mass differences (in MeV) 
calculated from the $b m_l=0.007$ and
$0.010$ MILC lattices plotted vs the pion mass composed of sea quarks.
Various data have been displaced 
horizontally by small amounts for display purposes. 
A lattice spacing of $b=0.125~{\rm fm}$ has been used to set the scale.
}
\label{fig:pqdata}
\vskip .2in
\end{figure}

Each pair of partially-quenched propagators generated four different proton
states and hence three different mass-splittings from the isospin symmetric QCD
proton state.
The results of our calculations are shown in fig.~\ref{fig:pqdata}.
By fitting four parameters, $\overline{\alpha}, \overline{\beta}, g_1$ and
$g_{\Delta N}$ to the six mass-splitting we are able to extract the combination
of constants in eq.~\ref{eq:mnminusmp}.
Further, using the ratio of light-quark masses determined by the MILC
collaboration, $m_u/m_d = 0.43\pm 0.01\pm 0.08$~\cite{Aubin:2004fs}, we can
make a prediction for the strong-isospin breaking contribution to the
neutron-proton mass-splitting of
\begin{eqnarray}
M_n-M_p\big|^{d-u} & = & 2.26\pm 0.57 \pm 0.42 \pm 0.10~{\rm MeV}
\ \ \ ,
\label{eq:looplevelnumbers}
\end{eqnarray}
where the last error is an estimate of the systematic error due to
truncation of the chiral expansion.
This is to be compared with a value of 
$M_n-M_p\big|^{d-u} = 2.05\mp 0.30~{\rm MeV}$ as determined via the Cottingham
sum-rule.

\section{Conclusions}
\label{sec:resdisc}

We have performed partially-quenched calculations of the proton mass with
domain-wall valence quarks on the isospin-symmetric coarse staggered MILC lattices~\cite{Beane:2006fk}.
These calculations have allowed us to isolate the contribution to the
neutron-proton mass-splitting due to the light-quark mass-splitting at leading
and next-to-leading order in the 
$SU(2)_L\otimes SU(2)_R$ chiral expansion.
We find a value that is consistent with the estimate arrived at
from the experimental mass-splitting and the best estimate of the
electromagnetic contribution.  It is clear that with more computer power, this
important quantity will be calculated to high precision.

\end{document}